%% file: main.tex
\documentclass[fleqn,10pt]{wlscirep}
\usepackage[utf8]{inputenc}
\usepackage[T1]{fontenc}
\usepackage{lineno}

\usepackage{multirow}
\usepackage{makecell}
\usepackage{nameref}
\usepackage{hyperref}
\usepackage{booktabs,tabularx}
\usepackage{subcaption}
\usepackage{tabularx}
\usepackage{ragged2e}
\usepackage{array}
\definecolor{darkred}{RGB}{222,0,0}
\newcommand{\rev}[1]{\textcolor{black}{#1}}

\newcommand{\dataset}{\texttt{GazeDepth}~}

\title{An Eye-Tracking Dataset for Viewing Distance \rev{Categories} in Real-World Scenarios}

\author[1,*]{Dohwa Kim}
\author[1,*]{Yejin Choi}
\author[1]{Seungbok Lee}
\author[2]{Chi Yoon Jeong}
\author[1,$\dagger$]{Eunji Park}

\affil[1]{Chung-Ang University, Department of Computer Science and Engineering, Seoul, 06974, South Korea}
\affil[2]{Electronics and Telecommunications Research Institute, Human Sensory Augmentation Research Section, Daejeon, 34129, South Korea}

\affil[*]{These authors contributed equally to this work: Dohwa Kim and Yejin Choi.}
\affil[$\dagger$]{Corresponding author: Eunji Park (eunjipark@cau.ac.kr)}

\begin{abstract} 
Estimating viewing distance from gaze behavior is essential for understanding user intent and enabling distance-aware interactive systems.
However, most existing eye-tracking datasets have been collected in constrained settings, such as laboratory environments or static tasks.
Consequently, they only partially capture viewing behaviors in real-world situations where viewing distance changes with natural head and body movements.
We introduce GazeDepth, an eye-tracking dataset collected from 19 participants using a wearable tracker during tasks reflecting real-world scenarios. GazeDepth includes \textit{fixed-distance viewing scenarios} with constant observer–target distances at near (33 cm), middle (50 cm), and far (300 cm), as well as \textit{variable-distance viewing scenarios} in which participants shift gaze among targets at different depths in indoor and outdoor environments. The dataset provides synchronized gaze data, pupil size, 3D eye-vectors, and head-motion signals, along with distance labels. 
\rev{Statistical analyses showed that distance-related gaze features, such as vergence angle and estimated viewing distance, differed consistently across viewing-distance categories. In addition, classification models trained on GazeDepth further demonstrated that the dataset captures gaze characteristics that distinguish the three viewing-distance categories, supporting gaze-based distance inference and distance-aware interaction in realistic scenarios.}
\end{abstract}

\begin{document}
\flushbottom
\maketitle
\thispagestyle{empty}
\section{Background \& Summary}
\rev{
Visual activity provides crucial cues for inferring human intentions and cognitive states, as vision accounts for more than 80\% of the sensory information processed by the human brain~\cite{qiu2024vision}.
For example, longer fixations and changes in saccade patterns can indicate increased difficulty in processing visual information~\cite{liversedge2000saccadic}, while gaze toward objects can provide cues about which object a person is likely to interact with next~\cite{huang2015using}.}
Accordingly, there is a strong need for reliable and scalable methods to quantitatively measure visual behavior, and eye-tracking technologies have attracted significant attention in research~\cite{duchowski2002breadth,pluzyczka2018first}.
Eye tracking refers to techniques that estimate the point of gaze and measure eye movements (e.g., fixations and saccades) by analyzing video images of the eye, or by using physiological sensors such as electrooculography (EOG)~\cite{duchowski2002breadth,guestrin2006general}.


\rev{
Most prior gaze-estimation research has focused on predicting gaze positions on two-dimensional planes, such as screens and other planar surfaces. These approaches have achieved high accuracy and enabled applications in domains including marketing, assistive technology, and medical diagnosis~\cite{wedel2017review,majaranta2002twenty,bigolin2022reflacx}. 
However, two-dimensional gaze positions alone may be insufficient to identify what a user is attending to in real-world environments, where multiple objects can lie along similar viewing directions but at different depths. Gaze depth is therefore particularly important in scenarios where distinguishing among objects located at different distances is necessary. Estimating gaze depth addresses this limitation by providing information about how far the attended target is from the user.
This information is particularly valuable in applications where system behavior depends on the viewing distance of the attended target~\cite{mansouryar20163d, mardanbegi2019resolving, zhang2024focusflow}. For example, in automotive head-up displays, gaze depth can help determine whether a driver is looking at a nearby UI element or monitoring the distant road~\cite{seraj2024realtime}. In head-worn wearable displays, such as smart glasses, viewing distance information can support the appropriate presentation of information as the user's viewing distance changes~\cite{dunn2019required}.
}

Prior studies have investigated how gaze characteristics vary with viewing distance and whether these characteristics can be used for gaze depth estimation~\cite{toates1974vergence,riggs1960eye,feil2017interaction}.
However, most studies investigating these cues have been conducted in constrained environments (see Table~\ref{tab:distance_datasets}).
For example, many existing studies have been conducted in settings where the user’s head was fixed, and the targets remained stationary in front of the user. This setup is far from real-world conditions, in which viewing distance changes dynamically with natural head and body movements~\cite{kwon20063d, kuo2018depth}.
This may limit their ability to capture distance-specific gaze characteristics in natural viewing contexts.
Other studies were conducted in VR environments that differ from real-world viewing conditions and may involve display-specific confounding factors such as the vergence–accommodation conflict~\cite{hoffman2008vergence,adhanom2023eye,hepperle2023similarities}.


\rev{
To address this gap, we developed \dataset, an eye-tracking dataset designed to capture}
gaze behavior across multiple viewing distance \rev{categories} in tasks reflecting real-world scenarios.
To capture gaze behavior under both fixed and time-varying viewing distances, which may elicit different depth-related cues over time, \rev{we organized the experimental protocol into a set of baseline measurements and two main task types}: \textit{fixed-distance viewing scenarios} and \textit{variable-distance viewing scenarios}.
\rev{
Based on distance criteria adopted in prior multifocal lens design studies, we set three observer–target distance conditions: near (33 cm), middle (50 cm), and far (300 cm)~\cite{wolffsohn2024bcla,hogarty2018comparing,mojzis2014comparative,alfonso2016visual}.
The baseline measurements consisted of dot fixation, surface viewing, and multi-object viewing and were designed to characterize individual differences and establish reference gaze patterns.}
\rev{
In the fixed-distance viewing scenarios, participants performed book-reading, web-browsing, and object-seeking tasks at each distance category. 
These tasks provided controlled viewing-distance conditions while allowing natural head and body movements, complemented by the more dynamic variable-distance scenarios.
In the variable-distance viewing scenarios, participants performed an indoor board-game and an outdoor wayfinding task, both of which elicited natural gaze shifts across near-, middle-, and far-distance target categories.
}

\dataset includes fixation, saccade, and blink events, along with pupil size, 3D gaze direction, eye position vectors, and head Inertial Measurement Unit (IMU) data, recorded using the Pupil Labs Neon eye tracker. It comprises a total of 22.9 hours of recordings from 19 participants and contains 213,881 fixation events.
To verify that the dataset captures distinct gaze characteristics at different viewing distance \rev{categories}, we conducted statistical analyses using gaze-related indicators known to vary with viewing distance (e.g., pupil size and vergence angle), as proposed in previous studies~\cite{toates1974vergence,riggs1960eye,feil2017interaction}.
\rev{
The results revealed significant differences across near, middle, and far conditions, indicating that \dataset reliably captures distance-dependent variations in gaze behavior.
We also conducted machine-learning classification using gaze features to evaluate whether \dataset supports subject-independent classification of the three viewing-distance categories, following prior work of machine learning to gaze analysis under natural-viewing conditions~\cite{nejad2024ace}. 
We compared standard and widely used six classifiers: a multilayer perceptron (MLP)~\cite{rumelhart1986learning}, decision tree (DT)~\cite{loh2011classification}, Random Forest (RF)~\cite{breiman2001random}, Linear Discriminant Analysis (LDA)~\cite{balakrishnama1998linear}, XGBoost~\cite{chen2015xgboost}, and k-nearest neighbors (kNN)~\cite{duda1973pattern}.
}
LDA achieved the highest accuracy in the fixed-distance viewing task (0.966), while RF performed best in the variable-distance tasks, reaching accuracies of 0.827 (indoor) and 0.763 (outdoor).
These classification performance further show that \dataset contains meaningful signals for learning to discriminate viewing distance \rev{categories}.

\input{Table/tab_other_research}

\section{Methods}
\subsection{Data Collection Protocol} 
\input{Figure/fig_study_protocol}
The data collection protocol was designed with two considerations: (1) to elicit measurable differences in gaze characteristics and gaze transition patterns across distinct distance conditions, and (2) to provide viewing scenarios that resemble real-world situations. These design choices enable the study of gaze dynamics underlying natural visual attention and gaze-shift processes.

Data collection comprised baseline measurements and five main tasks, organized into two task scenarios: Fixed-distance viewing scenarios (three subtasks) and Variable-distance viewing scenarios (two subtasks).
The baseline measurement was conducted under a relaxed viewing condition to capture individual differences in gaze characteristics. 
\rev{It consisted of three subtasks: (1) a dot-fixating task (Figure~\ref{fig:figure1}-a), which assessed each user's concentration and oculomotor stability within a brief period~\cite{ogle1959depth, green1980depth, annerer2021reliably}, (2) a surface-viewing task (Figure~\ref{fig:figure1}-b), which measured dwell gaze characteristics under minimal visual stimulation~\cite{holmqvist2011eye}, and (3) a multi-object viewing task (Figure~\ref{fig:figure1}-c), which examined gaze distribution and transition patterns in complex visual scenes~\cite{deng2024advancing}.}

The main experiment consisted of two scenario types: (1) Fixed-distance viewing scenarios (Figure~\ref{fig:figure1}-d, e, f), in which the relative distance between participants and visual targets remained constant at near, middle, and far positions, and (2) Variable-distance viewing scenarios (Figure~\ref{fig:figure1}-g, h), in which participants frequently shifted their gaze \rev{across near-, middle-, and far-distance target categories~\cite{foulsham2015eye}.}
\rev{The fixed-distance viewing scenarios aimed to compare gaze characteristics across near-, middle-, and far-distance categories in realistic task contexts, including book reading~\cite{rayner1998eye, cop2017presenting}, monitor-based browsing~\cite{manhartsberger2005eye}, and screen-based visual target search~\cite{zelinsky2008eye, hwang2011semantic}, respectively.
The variable-distance viewing scenarios elicited gaze transitions in real-world indoor and outdoor environments, enabling analyses of gaze patterns during transitions across viewing distance categories.
It included an indoor board-game task and an outdoor wayfinding task~\cite{burch2017visual, wunderlich2021eye}.
The outdoor wayfinding task further captured gaze data under real-world environmental disturbances, such as illumination changes and wind, allowing examination of their potential effects on gaze patterns.
Detailed task procedures are described in Section~\ref{sec:data_acquisition_procedure}.}

\subsection{Ethics Statement} 
The construction of the \dataset was approved by the Institutional Review Board (IRB) of Chung-Ang University. The Chung-Ang University Ethics Center reviewed and approved the consent form, which included the purpose of data collection, procedures, types of data to be collected, data protection protocols, and compensation for participants (IRB No. 1041078-20240819-HR-225). Before participation, participants were given a consent form and asked to read it carefully. All participants provided written consent to data collection. Participants were informed that they could withdraw from the study at any time.

\subsection{Participants}
\input{Table/tab_demography}

A total of 19 participants were recruited through public advertisement posts on a university community board and a local information-sharing platform. Eligible participants were adults aged 19 years or older with no ophthalmic disorders and no difficulty in object fixation or walking during data collection. The recruited participants consisted of 9 aged 20–39 years and 10 aged 40–59 years (mean age = 39.21, \textit{SD} = 11.83). Since the eye tracker supported attachable lens values from -3.0 to +3.0 diopters, participants included individuals with corrected visual acuity of 1.0 or higher, those within the -3.0 to +3.0 diopter range, and contact lens wearers whose vision was corrected to their normal acuity. 

The recruitment notice informed participants of potential mild physical discomfort and a session duration of up to 2 hours. Participants were also required to remove any makeup around the eyes and wear a glasses-type eye tracker.
Participants were also informed that the tasks involve simple, non-strenuous everyday visual activities, and that all collected data would be used solely for research purposes. Each participant received compensation of approximately \$14 for participation.
Detailed participant information is provided in Table~\ref{tab:participant_demographics}. 

\subsection{Data Collection Setup} 
\input{Figure/fig_setup}
Data collection for indoor tasks was conducted in two classrooms under controlled illumination. For the baseline measurements and fixed-distance viewing scenarios, the distance between each participant and the target objects was fixed in advance to maintain consistent viewing distances.
Additionally, items prepared for data collection included books, mouse and keyboard, the “Kushi Express~\cite{mandoogames_publishing_kushi_express}” board game, a 27-inch monitor and 118-inch screen, a Pupil Labs Neon eye tracker ("I can see clearly now" frame)~\cite{baumann2023neon}, and a neck strap for carrying a smartphone (motorola edge 40 pro) while walking. 

Participants wore an eye tracker connected to a smartphone during data collection (see Figure~\ref{fig:setup}). If necessary, corrective lenses were attached to the eye tracker. The experiment was conducted after thoroughly confirming that the eye tracker did not obstruct the participant's field of view. The data collected from the eye tracker were transferred to a connected mobile device and then automatically uploaded to Pupil Cloud, a cloud-based storage system. Within the Pupil Cloud, gaze events such as fixations and saccades were automatically detected, labeled, and stored as part of the dataset.

\subsection{Data Acquisition Procedure}\label{sec:data_acquisition_procedure}
Data collection sessions were conducted in four stages: (1) onboarding, (2) baseline measurement, (3) fixed-distance viewing scenarios, and (4) variable-distance viewing scenarios (see Figure~\ref{fig:figure1}). A two-minute break was provided between each task session, and when a location change was required, participants removed and reattached the eye tracker. Two experimenters conducted all data collection, manually starting and stopping the eye-tracker recordings.

Additionally, although the Pupil Labs Neon eye tracker is an end-to-end deep learning-based calibration-free device that does not require a separate calibration procedure, we performed gaze offset correction using the post-offset correction method provided by Pupil Cloud to enhance data accuracy~\cite{pupil_labs_offset_correction}. Since recording was conducted independently for each task, a calibration video was also recorded separately before each task to enable task-specific post-processing correction. For the variable-distance outdoor scenarios, continuous device movement during walking made it difficult to maintain a stable initial calibration state. Therefore, data were directly collected using the Neon eye tracker's native gaze estimation pipeline without performing an additional initial calibration step.

\subsubsection{Onboarding}
Before data collection, participants completed a pre-survey that gathered demographic information (e.g., gender, age, visual acuity, ocular disease, and glasses/contact lens use) and task-related habits (e.g., focus while reading and difficulties when browsing digital content). This survey confirmed that participants could perform the tasks without any difficulty, and responses were fully anonymized using participant IDs.
Upon arrival at the experimental location, participants received research instructions and completed a consent form. Participation was voluntary, and consent could be withdrawn at any time without disadvantages. Afterward, participants were briefed on the overall experimental process, the eye tracker use, and precautions. If necessary, attachable lenses matching their individual visual acuity were worn. Experimenters confirmed the eye tracker was securely fitted, did not obstruct the field of view, and recorded data properly before the session began.

\subsubsection{Baseline Measurement}
Participants conducted three baseline measurements about 10 minutes: dot-fixating, surface-viewing, and multi-object viewing.
\paragraph{Dot-fixating task.} Participants sequentially sat and fixated on a central dot of the monitor for at least 4 seconds~\cite{harezlak2014towards} at pre-assigned near (33 cm), middle (50 cm), and far (300 cm) positions. To maintain participant focus, the dot size expanded and contracted at a constant rate~\cite{krafka2016eye}, and was scaled by distance to ensure visibility. Specifically, the dot diameters were set to 2.6-5.2 cm at near, 4.9–9.8 cm at middle, and 7.6-15.2 cm at far positions. 
\paragraph{Surface-viewing task.} Participants freely gazed at a blank screen (60 $\times$ 32 cm) at the middle position (50 cm). They were pre-informed that the screen color would change from white (5s) to black (10s). 
\paragraph{Multi-object viewing task.} Participants freely viewed 50 multi-object images from the MS-COCO dataset~\cite{lin2014microsoft} and 7 images of the experiment environment at the middle position (50 cm). Each image was displayed for 5 seconds, preceded by a 2 seconds guidance text message before the next image appeared. Participants fixated on a central dot for calibration before starting.

\subsubsection{Fixed-distance viewing scenarios}
\paragraph{Near-distance viewing task.} Participants read specific pages of a given book for 10 minutes. Before starting, they fixated on a sailboat on the book cover that served as a salient target for calibration.
They then read naturally at their preferred posture, and the viewing distance for the book was approximately 25–35 cm. The reading material was a school magazine containing a mix of text and images.
\paragraph{Mid-distance viewing task.} Participants completed the following two subtasks using a keyboard and mouse on the monitor 50 cm ahead: (1) creating a new Google account, and (2) web browsing following the experimenter's audio instructions (e.g., “Check the library location” and “Verify the university's founding year”). 
For the account creation task, all required personal information (e.g., name and Google email address) was fictitious. 
For the web browsing task, mouse operation was restricted during audio instruction to clearly delineate task segments and to ensure that participants fully attended to and comprehended the instructions before proceeding. 
When the participant found the information according to the audio instructions, they were instructed to respond verbally.

All participants completed the task for at least 10 minutes, responding to 7-14 instructions based on individual pace. To mitigate bias associated with prior familiarity, two unfamiliar university websites were used as target pages.
Information on detail audio-instructed questions is provided with the dataset. To ensure that responses reflected active information search rather than pre-existing knowledge, only answers that were directly verified on-screen and correctly stated by the participant were accepted as correct. Participants fixated on a central dot for calibration before starting. 
\paragraph{Far-distance viewing task.} Participants identified visual targets within images or videos presented on a screen 300 cm ahead. Each text instruction (e.g., “Find the woman wearing black sunglasses” and “Count the number of words spelling ‘real estate’”) was followed by an image or video, and participants verbally reported the target’s number or location in detail to the experimenter. 10 images (up to 20s each), and 10 videos (up to 10s each, with a maximum of two replays) were presented. All stimuli were derived from Korean street scenes crawled from YouTube to recreate the visual scenery typically observed in everyday environments. Before starting, participants fixated on a central dot for calibration.

\subsubsection{Variable-distance viewing scenarios}
\paragraph{Variable-distance indoor task.} Participants played the board game “Kushi Express~\cite{mandoogames_publishing_kushi_express}”, which requires gaze transitions across different distances. In the game, participants made a skewer model consisting of one stick, two perforated cloth pieces, and four colored blocks with through-holes, and then rapidly placed it on a plate illustration. During this process, participants shifted their gaze among the manipulation object (near), the opponent player (middle), and the instruction screen (far).
Participants manipulated the game tools placed directly in front of them, with a near working distance of approximately 30–40 cm. The opponent player was positioned approximately 90–110 cm away, and the screen was placed approximately 300–350 cm away, corresponding to middle and far viewing distances.

Participants were instructed to sit on a chair at the desk where the board game was set up and received guidance on the game's rules. Participants who had never experienced this board game completed three practice rounds, then sequentially assembled 27 targets. They were instructed to complete their skewers quickly and accurately, continuing even if the opponent player finished first. They were also advised to observe the assembly process of the opponent player (a trained experimenter) during the experiment. The trained experimenter maintained a similar pace to the participants to induce competitiveness and focus. After each round, the experimenter verified the result and reset the materials to their initial positions. The experiment lasted at least 15 minutes. Participants performed a gaze calibration by fixating on a central dot on the screen before starting.

\paragraph{Variable-distance outdoor task.} Before starting the task, participants received safety guidelines (e.g., no phone use while walking and no road crossing) and wore an eye tracker connected to a smartphone device, which was placed in a phone-holding necklace. They then entered the experimental chat room via their own phones, which provided a designated route map with the route marked in red and destination building name. Two experimenters accompanied each participant to ensure safety, one following from a close distance and the other from farther back to monitor and control potential hazards along the route. Upon reaching specific locations, the experimenter provided searching tasks via messenger (e.g., “Find the phrases written on specific building walls” and “Identify the bus stop name”). Participants alternated gaze among the mobile phone (near), signage or display stands (middle), and buildings (far) while moving and confirming the instructed target, then replied via chat. The corresponding viewing distances were approximately 25-35 cm for near, 80-150 cm for middle, and 20 m or more for far. 

The experimenter confirmed the response and verbally instructed the participant to proceed. If participants attempted to deviate significantly from the route or became lost, the experimenter corrected them with minimal intervention. Each participant completed 5 exploration tasks along a route approximately 550 meters, lasting at least 10 minutes.

\subsubsection{Post Survey}
After completing all tasks, participants completed a post-survey. This survey assessed self-reported concentration (7-point Likert scale), the objects most frequently fixated, and the perceived similarity between the experimental and real environments. Responses were required to meet a minimum length to encourage thoughtful feedback. Questionnaire data are provided with the dataset.

\subsection{Viewing Distance Annotation}
Unlike fixed-distance tasks, in which the viewing distance is predetermined, the variable-distance viewing task requires labeling the viewing distance \rev{categories} for the participant’s eye-tracking data.
Therefore, we devised a reliable automatic distance labeling method for the variable-distance viewing task to facilitate the modeling and evaluation of viewing distance \rev{category} estimation and to increase the practical utility of the \dataset.

\subsubsection{Automatic Annotation for the Variable-distance Indoor Task}
\paragraph{Annotation algorithm.}
\input{Figure/fig_auto_labeling_indoor_sample}
To obtain \rev{category labels for viewing distance} in the variable-distance indoor task, we first divided front-facing scene videos into individual frames. 
We then adopted a two-step approach (see Figure~\ref{fig:auto_labeling_indoor_sample}): (1) dividing each frame image into three spatial zones (near, middle, and far), and (2) assigning a viewing distance \rev{category label to each frame based on which of these predefined spatial zones contained the gaze coordinate.}
To define the three \rev{zones}, we used the objects that consistently appeared across experimental sessions: the plate illustration (located between the near and middle objects), the opponent player (designated as a middle-distance object), and the table.
To obtain these object locations, we performed object segmentation with the following text prompts using Grounded-SAM-2~\cite{ren2024grounded}: “\textit{hand}. \textit{lady} (opponent player). \textit{screen}. \textit{plate}. \textit{brown table}.”
Additional object prompts were added to suppress misclassifications (e.g., preventing the participant's \textit{hand} from being mistaken for \textit{lady}, ignoring the \textit{plate} displayed on the \textit{screen} as part of the task instructions).

We then drew two lines to separate the \rev{zones}: (1) a line through the center of the \textit{plate} and (2) a line through the top center of the \textit{lady}. To correct camera distortion and the subject’s head pose, we calculated the slope of the reference lines by fitting a polynomial regression \rev{line to the pixel coordinates of the segmented \textit{brown table} mask.}
The color brown was specified to enhance detection accuracy, as the experimental table was brown.

The near \rev{zone} was defined as the area below the \textit{plate}, the middle \rev{zone} as the area between the \textit{plate} and the \textit{lady}, and the far \rev{zone} as the area above the \textit{lady}.
In the middle \rev{zone}, we set an additional condition that the zone should be bounded by the x-coordinates of the \textit{lady} object, as the area beyond those limits often corresponded to the walls of the experimental space.
Finally, based on the classified spatial \rev{zones}, we assigned the viewing distance \rev{category} label to each video frame by temporally aligning the frame with the gaze coordinates recorded by the Pupil Labs Neon eye tracker that were nearest in time to the frame. 

\paragraph{Exception handling.}
\input{Table/tab_auto_scenario}
Four labeling cases occurred based on whether the \textit{plate} and \textit{lady} were detected (see Table~\ref{tab:case_definition_before_after}).
Across all participants, a total of 456,325 frames captured during periods when the task was actually performed were used for annotation. 
Within these frames, both reference objects (\textit{lady} and \textit{plate}) were detected in 81.36\% (Case 1), followed by \textit{lady}-only detections in 17.73\% (Case 2).
Frames where only the \textit{plate} was detected (Case 3) or no object was detected (Case 4) each accounted for less than 1\% of frames.

Given its relatively high frequency, we focused on Case 2. Case 2 reflected two possible situations: (1) the \textit{plate} was not actually present in the scene, or (2) the \textit{plate} was present but not detected.
To address missed \textit{plate} detections, we devised a method to estimate its position, considering the experimental setup that participants always faced forward.
Specifically, for each subject, we first computed the average distance between the \textit{lady} and the \textit{plate} across all frames in which both the \textit{lady} and the \textit{plate} are detected. We then estimated the position of the \textit{plate} by shifting the position of the \textit{lady} by the previously calculated distance along the perpendicular direction of the slope derived from the detected table.
If the estimated position extended beyond the scene boundary, this indicated that the subject was looking upward and only the \textit{lady} was visible, suggesting that the \textit{plate} lay outside the field of view (i.e., not present in the scene).
Using this approach, the \textit{plate} position was estimated for 18,913 frames (23.37\% of Case 2).

\paragraph{Annotation results.}
As a result, approximately 59.30\% (\textit{SD} = 10.48\%) of all frames were labeled as near, 16.59\% (\textit{SD} = 7.36\%) as middle, 21.45\% (\textit{SD} = 6.62\%) as far, and 2.66\% (\textit{SD} = 3.02\%) as unassigned (no matching target) within all frames.

\subsubsection{Automatic Annotation for the Variable-distance Outdoor Task}
\paragraph{Annotation algorithm.}
\rev{Unlike the variable-distance indoor task, where similar scene configurations repeatedly appeared across frames, the outdoor task involved more variable scenes and did not provide clear objects that could serve as stable boundaries for dividing the scene into distance zones. Therefore,} we divided front-facing scene videos into individual frames and assigned viewing distance \rev{category} labels by matching gaze coordinates to bounding boxes of distance-specific objects.
Specifically, we used Grounding DINO~\cite{liu2024grounding} to detect the following distance-specific objects predefined in our experimental setup in each frame: “near - \textit{cell phone}. \textit{hand}, middle - \textit{sign}. \textit{bus stop sign}. \textit{magazine rack}. \textit{brochure stand}. \textit{display stand}, far - \textit{building}." 
When bounding boxes overlapped (e.g., a bus stop sign in front of a building) and multiple objects corresponded to the gaze coordinates, we selected the smaller bounding box (i.e., the foreground object) as the matched target.

\paragraph{Annotation results.}
As a result, approximately 30.15\% (\textit{SD} = 3.89\%) of all frames were labeled as near, 10.97\% (\textit{SD} = 2.18\%) as middle, 24.66\% (\textit{SD} = 1.94\%) as far, and 34.21\% (\textit{SD} = 3.00\%) as unassigned (no matching target) within an average of 24,794.67 (\textit{SD} = 4,173.07) frames per participant.
For the outdoor task, as locomotion periods were considered part of the task along with the searching period, we included all frames recorded during the entire task duration.
As participants walked through real-world environments during the outdoor task, they frequently looked at task-irrelevant objects, resulting in a higher proportion of unassigned labels.

\subsubsection{Manual Annotation for Automatic Annotation Validation}
\input{Figure/fig_annotator_tool}
To validate the proposed automatic annotation method, manual annotations were conducted on videos recorded during the variable-distance indoor and outdoor tasks. Videos from six participants were randomly selected for each task without overlap (indoor: P1, 9, 11, 13, 15, 19, outdoor: P4, 5, 7, 10, 12, 18). Each task was independently annotated by a separate group of three annotators (a total of six annotators), and within each task, all selected videos were annotated by all three annotators.
Manual annotation was performed only on fixation events \rev{detected by the Pupil Labs fixation detector, as described in Section~\ref{sec:data_records_eyetracking_data}.}

For each fixation event, the central frame was annotated as a stable representative gaze state, and the assigned label was propagated to the entire fixation interval.
Annotators were instructed to assign labels based on the object in the image where the fixation coordinates were displayed, which were mapped to viewing distance categories as defined in the experimental design: indoor - participants’ hands or tools (near), experimenter or their tools (middle), and the display screen (far); outdoor -  participants’ phone or hands (near), signage or display stands (middle), and buildings (far). 
Annotators were instructed to annotate using accurate and consistent criteria.

Specifically, annotators were provided with an annotation tool as shown in Figure~\ref{fig:annotator_tool_indoor_outdoor}
and instructed to input numeric keys 0-4. 1, 2, 3, 4, and 0 corresponded to \textit{near}, \textit{middle}, \textit{far}, \textit{none} (not applicable), and \textit{skip} (indeterminable), respectively, and annotators were instructed to avoid using \textit{skip} whenever possible. Detailed object categories were provided in advance, and the tool displayed hints for each number, a progress bar, and a log of the last five entries. If an incorrect label was entered by mistake, the left arrow key allowed returning to the previous image. Images were automatically saved after each subject was completed, with manual saving possible at any time. 
Annotators were recruited through an online community, and each annotator labeled approximately 12,000-14,000 fixation instances over about 2.5 hours, receiving compensation of approximately \$20.

\section{Data Records}
This section describes the dataset directory and file structure.
The \dataset folder consists of four subfolders: Baseline, Fixed-distance\_viewing, Variable-distance\_viewing, and TaskResult, which are organized according to the task categories described in the previous section.
The Baseline folder contains two subfolders (Dot-fixating\_Surface-viewing and Multi-object\_viewing). The Fixed-distance\_viewing folder is divided into Near-distance\_viewing, Mid-distance\_viewing, and Far-distance\_viewing subfolders, while the Variable-distance\_viewing folder is split into Variable-distance\_indoor and Variable-distance\_outdoor subfolders.
The TaskResult folder provides CSV files containing participants’ pre-/post-survey responses, task instructions, and subtask results.
All task folders except TaskResult include a Timeseries\_Data\_and\_Scene\_Video directory by default, along with additional subfolders for markers and annotations.
Task marker folders are provided for the multi-object viewing task, mid- and far-distance fixed-viewing tasks, and store the screen coordinates of the markers used in each task.
The distance\_labeling folder is provided only for the variable-distance viewing scenarios and contains labels indicating which target distance the participant was looking at.
Within Timeseries\_Data\_and\_Scene\_Video folder, there are participant-specific recording folders following the naming convention: \texttt{\{TaskName\}\_\allowbreak\{participantID\}\_\allowbreak\{date\}\_\allowbreak\{recordingStartTime\}\_\allowbreak\{uniqueID\}}.
\rev{Detailed descriptions of the file formats, data fields, and recording exceptions are provided in Supplementary Information Section 1.}

\subsection{Eye-Tracking Data}\label{sec:data_records_eyetracking_data}
Within each task’s Timeseries\_Data\_and\_Scene\_Video folder, there is a participant data folder and a \texttt{sections.csv} file.
Each recording folder contains an mp4 front-facing scene video capturing the participant’s first-person view during the task(i.e., \{beginning of sectionID\}\_\{sectionStartTime\}\-\{sectionEndTime\}.mp4), two meta JSON files, and eight CSV data files.
\rev{
The CSV files contain scene-video timestamps, task events, gaze coordinates, fixation, saccade, and blink events, binocular 3D eye states and pupil sizes, and IMU-based head-motion data. Records across the files can be linked using recording and section identifiers.
}
The scene videos were recorded at 60 fps with a resolution of 1600 $\times$ 1200 pixels, while both gaze data and IMU signals were sampled at 200 Hz.
In the shared videos, the faces of individuals other than the research team experimenters are anonymized. Experimenters who may appear in the recordings are members of the research team and have provided consent for their appearance, and no additional sensitive personal information is disclosed. Also, license plates visible in the outdoor task videos were anonymized by applying mosaic blurring.

\rev{Fixation and saccade events were automatically detected in Pupil Cloud v7.2 and stored in \texttt{fixations.csv} and \texttt{saccades.csv}, respectively.
Fixation events were defined as a period during which gaze is stabilized toward a visual target, including compensatory eye movements during head or body motion~\cite{hessels2018eye, drews2024strategies, pupil_labs_neon_data_streams}. They were detected using the Pupil Labs fixation detector, which extends the classic Identification by Velocity Threshold (I-VT) algorithm that classifies low-velocity gaze samples as fixations. To account for head- and body-motion effects, the Neon fixation detector uses gaze velocity after correcting for scene motion estimated from IMU data.
Further algorithmic details, including parameter settings, are provided in the Pupil Labs fixation detector algorithm documentation version 3.6.24~\cite{pupil_labs_neon_data_streams}.
Saccade events were derived from gaps between fixations~\cite{pupil_labs_neon_data_streams}.}

\subsection{Marker Mapper Data \& Screen Recording}
In tasks with recorded screen videos (i.e., the Multi-object viewing task, Mid-distance viewing task, and Far-distance viewing task), markers were attached to the four corners of the screen, and the eye tracker’s marker mapper enrichment was used to transform gaze coordinates into the screen coordinate system. By combining these transformed coordinates with the screen recording, the participant’s point of gaze on the screen can be estimated.
Within each of these three task folders, ScreenRecording subfolder and \{TaskName\}\_Task\_Marker subfolder are provided.
\rev{
These folders contain the screen-recording videos (i.e., \texttt{P\{n\}\_\{TaskName\}\_screenrecording}), a reference image (\texttt{reference\_image.jpeg}) used for surface mapping, and CSV files containing task-section information, detected screen-surface positions, and gaze and fixation coordinates projected onto the mapped screen surface.
}
For the Multi-object viewing task, a single screen recording is shared across the task.

\subsection{viewing distance Labels}
We provide the viewing distance labels described in \textit{viewing distance Annotation} section.
Within the Variable-distance\_viewing directory, we provide two subfolders, Indoor\_viewing\_distance\_labeling and Outdoor\_viewing\_distance\_labeling. 
Each subfolder contains each participant's distance labeling results collected in the corresponding environment (i.e., P{n}\_indoor\_labels and P{n}\_outdoor\_labels).
\rev{
These files contain automatically generated near-, middle-, and far-distance labels aligned with the scene-video frames and, where available, manual annotations and majority-vote labels used for validation.
}


\subsection{Task Result Files}
We provide xlsx files that contain participants’ pre-/post-survey responses, task instructions, and subtask results from the Mid-distance viewing, Far-distance viewing, and Variable-distance indoor/outdoor tasks.
The pre-/post-survey CSVs include the participant identifier (\texttt{subject\_id}), the original questionnaire items, and the corresponding participant responses.
The questionnaire items consist of either short-answer questions or 7-point Likert-scale questions. For Likert-scale items, the question text includes the anchors, specifying what responses 1 and 7 represent.
The task-result CSV includes \texttt{subject\_id} rows and columns in the format task\{n\}\_success for each subtask. For tasks where the instructions are needed (e.g., the mid/far/variable-outdoor tasks), the questionnaire items are provided in the first row. Subtask outcomes are recorded as O/X to indicate success or failure. When a subtask requires recording the participant’s response itself (e.g., counting the number of people in the far task), we provide the participant’s response.


\section{Technical Validation}
\subsection{Gaze Data Quality Validation}
We evaluated whether the gaze data in our collected dataset were recorded stably at 200 Hz sampling rate of the Neon eye tracker. We calculated the time interval between consecutive gaze samples from each recording and evaluated sampling-rate stability based on the expected 5 ms interval at 200 Hz.
Across all recordings, 0.18\% of timestamp intervals exceeded 5.5 ms. Among these exceeded intervals, the mean gap duration was 12.24 ms (SD = 12.53). The largest timestamp gap was 305 ms, which was observed near the beginning of a recording. This gap was likely caused by a brief sample drop during the initial stabilization of the recording stream, with minimal impact on overall sampling-rate stability.
At the participant level, the data loss rate was 0.18\% on average (SD = 0.06). The highest participant-level data loss rate was only 0.37\%. These results indicate that the gaze data were recorded stably at close to 200 Hz sampling rate, with only minimal sample loss.

\rev{We also calculated the angular gaze error after applying post-offset correction, using corrected gaze data and scene-camera video data.
The angular gaze error was defined as the angle between the two 3D rays in the camera coordinate system, one passing through the corrected gaze point and the other through the calibration target identified in the scene-camera frame~\cite{baumann2023neon}.
For each participant-task recording, angular gaze error was calculated from gaze samples within the fixation interval containing the calibration event, except for the gaze sample corresponding to the calibration event itself.
These 200 Hz gaze samples were matched to the calibration target identified in the 30 fps scene-camera frames using scene-frame timestamp bins.
To avoid gaze samples in which participants were likely not viewing the calibration target, such as those affected by brief head or scene-camera movements, angular-error values greater than four median absolute deviations from the within-fixation median were excluded~\cite{cuve2022validation, leys2013detecting}.
Angular gaze error was then summarized as the mean angular error for each participant-task recording, with values provided in Supplementary Table~S.5.
Across all recordings, the median angular gaze error was 0.70$^\circ$.
These results indicate that post-offset correction produced close alignment between the corrected gaze points and the calibration target.}

\rev{We further evaluated whether post-offset correction shifted gaze coordinates outside the valid 1600 $\times$ 1200 scene-camera image.
Across all recordings, only 0.05\% of gaze samples were outside the valid coordinate range.
At the participant level, the average out-of-frame rate was 0.05\% (\textit{SD} = 0.11), with the highest rate reaching 0.45\%.
The mean overflow distance from the image boundary was 26.84 px (\textit{SD} = 21.08). These results indicate that most corrected gaze coordinates remained within the scene-camera coordinate system.}

\subsection{Annotation Validation}
To validate the accuracy of the automated labeling method for both variable-distance indoor and outdoor tasks, we compared the automated label with manual annotations.
First, we assessed Krippendorff's $\alpha$~\cite{krippendorff2018content} which computes inter-annotator agreement for multiple annotators. Since viewing distance \rev{category} labels are nominal categories, we used nominal $\alpha$. Instances marked as \textit{skip} (i.e., difficult to judge) were treated as missing and excluded from the agreement calculation. According to the interpretation criteria for Krippendorff's $\alpha$, values above 0.80 indicate reliable agreement, values between 0.67 and 0.80 indicate moderate agreement, and values below 0.67 indicate insufficient reliability~\cite{krippendorff2018content}. All statistical results reported in this section include all data within the interval from \textit{task.start} to \textit{task.end}.

For the indoor task, a total of 14,604 frames were manually annotated (see \textit{Manual Annotation for Automatic Annotation Validation} section for further detail). On average, each participant's images comprised 2,434 frames (\textit{SD} = 588.20), and the participant-level mean $\alpha$ was 0.9556 (\textit{SD} = 0.0149). For the outdoor task, a total of 12,823 frames were manually annotated, with an average of 2,137.17 frames per participant (\textit{SD} = 342.92). The participant-level mean $\alpha$ was 0.7239 (\textit{SD} = 0.0566).
Although the agreement was lower due to greater diversity in objects and environmental conditions in outdoor scenes, it still reached a moderate level of agreement.
Detailed results are reported in \rev{Supplementary Table~S.3.}

We then compared the automated label with four manual label sets: three individual annotator labels and one majority vote label across annotators. The majority vote label was used as a consensus-based manual label, and tied majority vote cases were excluded from the analysis.
As described in \textit{viewing distance Annotation} section, the manual label assigned to the central frame of each fixation event was propagated to the entire fixation interval for comparison.
For each comparison, the analysis included only fixation frames for which a distance category (near, middle, or far) was assigned in the propagated manual labels. 
The mean and standard deviation of accuracy and \rev{macro} F1-score across the six participants' data are reported in \rev{Supplementary Table~S.4.}

In the variable-distance indoor task, compared with the annotators’ majority vote labels, the automated labels achieved an accuracy of 0.905 (\textit{SD} = 0.07) and an F1-score of 0.881 (\textit{SD} = 0.05) computed over an average of 24,079.5 frames (\textit{SD} = 3,490.49). In the variable-distance outdoor task, the automated labels achieved an accuracy of 0.822 (\textit{SD} = 0.05) and an F1-score of 0.81 (\textit{SD} = 0.05) against the annotators’ majority vote label, computed over an average of 11,610.83 frames (\textit{SD} = 2,932.72). These results indicate strong agreement between automated labels and consensus manual annotations across both tasks. 

\subsection{Distance-Dependent Gaze Signal Validation}
\input{Figure/fig_boxplot}

To validate that the gaze data in \dataset reliably capture differences in viewing distance \rev{categories}, we performed statistical analyses to examine their variation across distance conditions, and further conducted machine-learning and deep-learning based distance classification to assess the utility of \dataset.
We analyzed whether the eye-tracking features varied across viewing distance conditions (near, middle, and far) in three settings: (1) across the three fixed-distance tasks, (2) within the variable-distance indoor task based on the automatically assigned distance labels, and (3) within the variable-distance outdoor task based on automatically assigned distance labels.
\rev{Following prior work on viewing distance estimation~\cite{hosp2024simulation,arefin2022estimating}, we derived two distance-related features from the binocular eye-tracking data: the vergence angle and a geometry-based estimate of viewing distance. For each sample, the vergence angle was calculated as the inverse cosine of the dot product between the normalized left and right optical-axis vectors.
The two eyeball centers and the viewed target form a triangle, with the distance between the eyes as its base and the optical axes converging toward the target at the vergence angle~\cite{arefin2022estimating}. 
Based on this geometry, the viewing distance was estimated using the distance between the two eyeball centers and half of the vergence angle.
}
All measured and derived gaze features were z-score normalized within each \rev{participant-task session} to reduce inter-individual variability.

\subsubsection{Statistical Analysis}
We first assessed the normality of the gaze feature distributions for each distance condition using the Shapiro-Wilk test.
If the normality assumption was satisfied, we evaluated differences across the three conditions using repeated-measures ANOVA with paired t-tests for post hoc comparisons. Otherwise, we used the Friedman test as a non-parametric alternative for repeated measures, followed by Wilcoxon signed-rank tests for post hoc pairwise comparisons.
\rev{
Both vergence angle and the geometry-based viewing distance estimate differed significantly across the near, middle, and far conditions, with all pairwise comparisons reaching significance ($p<.01$) in all three settings (See Figure~\ref{fig:distance_validation_boxplots}).
Detailed statistical results for all features are provided in the Supplementary Table~S.6.
}

\input{Table/tab_modelResult}
\subsubsection{Distance-Dependent Vergence Characteristics}
\rev{
To further assess the usefulness of vergence angle for distinguishing viewing distances, following prior work showing its importance for viewing-distance estimation~\cite{arefin2022estimating, hosp2024simulation}, we conducted additional analyses. Specifically, we examined its distance-dependent characteristics and the time required to reach the mean vergence angle of a new viewing-distance condition.
Detailed results for these analyses are provided in Supplementary Information Section 5.
The mean raw vergence angles across participants were $13.11^{\circ}$ ($SD = 6.80^{\circ}$), $10.16^{\circ}$ ($SD = 6.50^{\circ}$), and $6.86^{\circ}$ ($SD = 4.92^{\circ}$) for the near, middle, and far conditions, respectively.
Using the geometric relationship between vergence angle and viewing distance, these mean angles corresponded to $36.18$~cm ($SD = 19.70$~cm), $56.73$~cm ($SD = 41.80$~cm), and $87.56$~cm ($SD = 65.16$~cm), respectively.
These geometry-derived estimates may differ from the predefined viewing distances, as they are not direct distance measurements.
Despite the overall decrease in vergence angle with increasing viewing distance, the absolute values varied substantially across participants, consistent with prior work~\cite{arefin2025measuring}. 
We therefore performed the following analyses separately for each participant.
}


\rev{
First, to assess the ability of vergence angle to distinguish adjacent viewing-distance categories, we estimated participant-specific decision boundaries for the near--middle and middle--far pairs.
For each pair, the decision boundary was defined as the vergence-angle threshold that maximized balanced accuracy~\cite{fluss2005estimation}.
The mean participant-specific thresholds were $11.37^{\circ}$ ($SD = 6.99^{\circ}$) for near--middle and $8.33^{\circ}$ ($SD = 5.91^{\circ}$) for middle--far.
We then measured how far the mean vergence angle of each distance condition was from the neighboring threshold. This indicates the amount of vergence change needed to move from one viewing-distance category to the next.
The mean differences were $1.75^{\circ}$ ($SD = 1.35^{\circ}$) for near-to-middle, $1.51^{\circ}$ ($SD = 0.82^{\circ}$) for middle-to-near, $2.38^{\circ}$ ($SD = 0.99^{\circ}$) for middle-to-far, and $1.75^{\circ}$ ($SD = 1.12^{\circ}$) for far-to-middle.}

\rev{
We also examined how quickly vergence adjusted after gaze shifted to a different viewing-distance category.
In the variable-distance indoor and outdoor tasks, each transition was identified by a change in the viewing-distance label.
When gaze shifted to a different viewing-distance category, a saccade occurred together with an adjustment in vergence toward the new distance condition. 
We characterized this adjustment using the saccade duration and the time required for vergence to adjust to the new viewing distance.
During viewing-distance transitions, saccades lasted an average of $148.5$~ms ($SD = 38.8$~ms), and vergence reached the mean angle of the new distance condition within $208.0$~ms ($SD = 53.4$~ms) on average.
From the end of the corresponding saccade, vergence reached the mean of the new distance segment after $139.3$~ms ($SD = 35.4$~ms) on average. 
These results indicate that vergence adjusted to changes in viewing distance within a few hundred milliseconds, suggesting that the $0.5$-s analysis window is sufficiently long to capture the vergence response following a change in viewing distance.
}

\rev{
These results demonstrate that vergence angle provides useful information for discriminating viewing distance categories. However, the substantial variability in participant-specific vergence values and decision boundaries suggests that applications relying on absolute vergence angle for individual distance classification may benefit from an initial calibration to establish user-specific distance distributions and decision boundaries.
}

\subsubsection{Viewing Distance Classification}
To evaluate whether \dataset supports viewing distance \rev{category} classification, we trained machine-learning classifiers and reported their classification performance.
Specifically, the classifiers included commonly used machine-learning models: DT~\cite{loh2011classification}, RF~\cite{breiman2001random}, kNN~\cite{duda1973pattern}, LDA~\cite{balakrishnama1998linear}, XGBoost~\cite{chen2015xgboost}, and MLP classifier~\cite{rumelhart1986learning} implemented using the scikit-learn library~\cite{scikit-learn}.
We set the hyperparameters of the machine learning classifiers (DT, RF, LDA, KNN, and XGBoost) based on settings in prior wearable-sensing studies~\cite{schmidt2018introducing, di2020multi, mirjafari2019differentiating}.
For the MLP classifier, we used the default scikit-learn settings following prior work on gaze depth estimation~\cite{lee2017estimating}.
We trained viewing distance \rev{category} classification models under the same three settings as the statistical analyses.
As inputs, we used the gaze signals recorded in \textit{3d\_eye\_states} along with the computed vergence angle and the vergence-based distance. 
\rev{All features were z-score normalized separately within each participant-task session using the full session recording.
We then segmented the normalized signals} using three window sizes (0.5 s, 1 s, and 10 s) with a step size equal to half of each window length (i.e., 50\% overlap).
We evaluated machine learning model performances using LOSO cross-validation to assess subject-independent generalization.
We report accuracy and macro F1-score averaged across folds as evaluation metrics. To ensure robustness of machine learning models, we repeated training five times and report the mean and standard deviation of the metrics.

\input{Figure/SHAP_all}

The classification results showed that the models could distinguish viewing distance conditions across all three settings.
In the fixed-distance viewing task, LDA achieved the highest accuracy of 0.966 ($SD = 0.04$) with a 10 s window. In the variable-distance viewing tasks, RF achieved the best performance with a 0.5 s window, reaching accuracies of 0.827 ($SD = 0.05$) in indoor task and 0.763 ($SD = 0.04$) in outdoor task.
These results suggest that our dataset contains meaningful signals for learning to distinguish viewing distance \rev{categories}.
\rev{These modeling results should be interpreted as an offline validation of viewing distance category discriminability, in which participant-task session statistics were available for feature normalization. For real-time or new-user applications, a brief calibration or adaptation step may be needed to derive normalization parameters before classification.}
The detailed modeling results are reported in Table~\ref{tab:model_performance_all}.
In addition, we conducted feature importance analyses using SHapley Additive ex-Planation (SHAP)~\cite{lundberg2017unified}, and the results are shown in Figure~\ref{fig:shap_all}.
\rev{The SHAP results showed that computed vergence angle and vergence-based viewing distance were important features for viewing-distance classification, consistent with prior work~\cite{toates1974vergence,riggs1960eye,feil2017interaction}. This supports the usefulness of vergence-related features for viewing-distance category estimation.
The optical-axis y feature also showed high importance, so we further examined the potential influence of vertical gaze direction through the ablation analyses in Section~\ref{sec:ablation}.}

\subsubsection{Ablation Study}
\label{sec:ablation}
\rev{
To further examine the contribution of individual gaze features and the influence of task-specific characteristics on viewing-distance classification, we conducted additional ablation and inter-task analyses.
First, we grouped the gaze features by feature type, including vergence-related (i.e, vergence angle and geometry-based distance), optical-axis, eyeball-center, and pupil-diameter features, and evaluated each group separately to quantify its contribution to classification performance.
When models were trained using each feature group individually, the optical-axis features achieved the highest mean accuracy of 0.799 ($SD = 0.05$), showing the smallest performance decrease relative to the full-feature model, followed by the vergence-related features with a mean accuracy of 0.706 ($SD = 0.09$).
Detailed results are provided in the Supplementary Information Section 6.
This pattern was consistent with the SHAP-based feature-importance analysis. 
Vergence-related features generally ranked among the most important features, together with the optical-axis \(y\) components that reflect vertical gaze direction.
The optical-axis y components may rank highly because they reflect vertical gaze direction, which can vary across viewing distance conditions depending on target location, such as the height of the target object.}

\rev{
To assess whether vertical gaze direction or binocular information was a major source of classification performance, we evaluated models with these features restricted.
We evaluated monocular conditions using only the left- or right-eye features and a condition in which features related to vertical gaze direction were excluded.
Across all tasks, the mean accuracy remained 0.785 ($SD = 0.07$) after excluding the optical-axis \(y\) features and reached 0.802 ($SD = 0.04$) when only the left-eye and right-eye features were used, respectively.
These results indicate that the classification performance did not rely predominantly on vertical gaze information or binocular information alone.}

\rev{
To examine whether viewing-distance classification relied on task-specific information, we conducted inter-task analyses.
Models were trained on the baseline dot-fixating task, in which viewing distance was systematically varied while other task-related factors were minimized, and then tested on the fixed-distance, indoor, and outdoor tasks using participant-level LOSO evaluation.
The resulting accuracies were 0.750 ($SD = 0.10$) for the fixed-distance task, 0.601 ($SD = 0.07$) for the variable-distance indoor task, and 0.555 ($SD = 0.05$) for the outdoor task. 
Although inter-task evaluation resulted in a mean accuracy decrease of 18.6 \% points relative to within-task evaluation, substantial classification performance was retained, and accuracy remained above the majority-class baseline in all three tasks.
These results indicate that the models captured viewing-distance-related gaze patterns that remained informative across different tasks, supporting that \dataset contains gaze information relevant to viewing distance beyond task-specific characteristics.
}

\section{Usage Notes}
The~\dataset~\cite{kim2025gazedepth} is available upon request via Zenodo~\href{https://doi.org/10.5281/zenodo.18625669}{https://doi.org/10.5281/zenodo.18625669}
The currently released version is distributed under a controlled access policy with a CC BY-NC-ND 4.0 license. This restriction reflects the inclusion of indirect identifiers or other sensitive information in certain data components (e.g., high-resolution gaze features and obfuscated images). Researchers seeking access to the dataset are required to follow the procedures specified on Zenodo and submit a data usage agreement (\dataset Dataset Data Use Agreement.pdf). The agreement must describe the intended research purpose, data storage and security measures, and a formal commitment to refrain from re-identification attempts. Requests are typically reviewed and processed within several working days. A detailed description of the data file structure and its individual components is provided in \textit{Data Record} section of this paper. 

When using the~\dataset, researchers should also consider the characteristics of the provided labels. The auto labels for the variable-distance viewing task in the dataset are discrete categories defined with respect to reference targets in the experimental design, rather than continuous measures of absolute distance. Therefore, these labels should not be interpreted as absolute distance without caution.

\subsection{Possible Applications}
First, our dataset can serve as a benchmark for evaluating models for distance classification, human behavior classification, and intention recognition in real-world environments. 
Unlike datasets collected in controlled laboratory settings, \dataset supports evaluation of model generalization across naturalistic gaze transitions at near, mid, and far distances.
For example, it can be used to evaluate whether a real-time distance estimation model can infer the currently attended target distance from gaze data.
It can also recognize an intention to shift focus from a nearby object to a farther target~\cite{lee2017estimating}. 
The outdoor route-finding task additionally makes the dataset suitable for benchmarking models of visual attention, spatial cognition, and environmental exploration during navigation~\cite{alinaghi2025decoding}. 

Such distance estimation models can support gaze-based AR/MR interfaces and context-aware assistance systems by indicating whether the user is inspecting nearby objects, exploring the workspace, or monitoring farther surroundings. 
This could allow smart-glasses-based assistants to adapt feedback, such as detailed object information or workspace-level guidance, to the user’s current visual context~\cite{moon2024smart}. 
Distance information could also be incorporated into gaze-based activity recognition models~\cite{bektacs2024gaze} to support finer-grained activity classification.

Also, our dataset can support vision-training validation systems that analyze whether users can stably shift visual focus across target distances or show delayed transitions at specific distances.
For example, the dataset could be used to design supporting metrics for verifying whether gaze changes follow target distance transitions during vergence training. This could support exercises such as the Brock string test and help detect difficulties in near-distance focusing~\cite{alvarez2010vision}. 
Thus, our dataset could be extended toward gaze-based validation tools for quantitatively monitoring progress in vision training and evaluating distance-specific focus adjustment characteristics~\cite{rovira2025objective}. 

\subsection{Limitations}
In the variable-distance outdoor task, participants moved through an open environment with a variety of objects and frequent scene changes. In many frames, the manual annotation reference objects were absent, did not match the gaze point, or were too distant to be reliably labeled. As a result, a larger proportion of samples did not meet the manual labeling criteria and were labeled as \textit{none}. 
\rev{The outdoor task also showed relatively lower annotation agreement than the indoor task. To further improve annotation reliability in the outdoor task, future work could provide additional frames before and after each fixation event, allowing annotators to better infer the gaze target from the temporal gaze context in complex outdoor scenes.}
In addition, since viewing distance is provided as categorical labels rather than actual continuous distance values, the dataset may be less suitable for estimating exact viewing distances.

\rev{Another limitation arises from the trade-off between providing realistic viewing scenarios and maintaining experimental control. 
To reflect realistic viewing scenarios, the fixed-distance conditions used different activities at each distance, which may have introduced task-related confounds. 
Although the fixed-distance tasks allowed natural head and body movements to better reflect everyday viewing behavior, this flexibility may also have introduced variation in gaze direction across distance conditions. In particular, the relatively high importance of optical-axis y features suggests that vertical gaze direction, potentially influenced by participants’ posture and target placement, may have contributed to the observed differences across viewing-distance categories.
To assess the extent to which these factors influenced the results, we conducted ablation and inter-task analyses. These analyses showed that viewing-distance-related gaze information remained informative even when gaze-direction-related cues were reduced.} 
\rev{}
Participant-related factors may have affected both generalizability and recording quality.
Gaze behavior may differ across age groups, but older adults are underrepresented in our dataset due to recruitment challenges.
Nevertheless, the study included participants with diverse visual conditions, providing some variability in ocular characteristics.
Additionally, because the eye tracker frame was not adjustable, mismatches between the device and individual head shapes may have affected wearing comfort and potentially influenced participants’ concentration during the task. Some participants occasionally adjusted the device while performing the task.

\section{Code Availability}
The code for processing this dataset (including automated annotation and anonymization) is accessible via GitHub~\url{https://github.com/ixlab-cau/GazeDepth}
.
All relevant details, including instructions for downloading the required model, are documented in the repository.

\clearpage

\bibliography{reference}

\section*{Acknowledgments}
The authors thank Yejin Jang for assisting with the experimental sessions and data collection.
This work was supported by Institute of Information \& communications Technology Planning \& Evaluation (IITP) grant funded by the Korea government(MSIT) (No.RS-2026-25514284, Development of technology to build a learning dataset for creating emotional AI models), and Electronics and Telecommunications Research Institute (ETRI) grant funded by Korean government [26ZR1200, Research on Autonomous Vision Augmentation and Extension Technologies]. Also, this research was supported by the Chung-Ang University Graduate Research Scholarship in 2025.

\section*{Author Contributions Statement}
D.K. and Y.C. contributed equally to this work. D.K. and E.P. conceptualized the study and designed the research. D.K. contributed to experimental sessions and data acquisition. D.K. and Y.C. contributed to data annotation and developed the annotation tools and pipelines, with D.K. focusing on outdoor annotation and anonymization and Y.C. focusing on indoor annotation and the annotation tool. D.K., Y.C., and S.L. verified and edited the dataset for anonymization and data preparation. D.K. and Y.C. conducted the data analysis and interpretation with support from S.L. D.K. focused on the statistical analysis and modeling, while Y.C. focused on the annotation validation. All authors conducted discussions and verification to reach consensus on the results. D.K. and Y.C. wrote the original draft with support from S.L. in designing the figures and tables. D.K., Y.C., S.L., C.J., and E.P. reviewed and edited the manuscript. E.P. oversighted of the planning and execution of research activities, including mentoring and funding acquisition. All authors reviewed the draft and final manuscript.

\section*{Competing Interests}
The authors declare no competing interests.

\clearpage

\end{document}

%% file: Table/tab_other_research.tex
\newcolumntype{Y}{>{\RaggedRight\arraybackslash}X}

\begin{table}[t]
\caption{\rev{Summary of eye-tracking datasets that provide viewing distance labels. For each dataset, we summarize experimental conditions (viewing distance, target condition, setting, and task type), whether distance estimation was performed in the study, number of participants, collected data types, and sensing devices. 
The datasets were collected in real-world (\textbf{Indoor} and \textbf{Outdoor}) and VR/AR/XR settings. 
The \textit{Distance estimation} column indicates whether the study conducted viewing-distance inference using the dataset.
The \textit{Viewing distance} column denotes a continuous range of viewing distance as \textbf{R} and discrete distances as \textbf{D}.}}
\centering
\scriptsize
\setlength{\tabcolsep}{3.5pt}
\renewcommand{\arraystretch}{1.25}

\begin{tabularx}{\linewidth}{
  >{\RaggedRight\arraybackslash}p{2cm}    
  >{\RaggedRight\arraybackslash}p{2cm}    
  >{\RaggedRight\arraybackslash}p{1.9cm}    
  >{\RaggedRight\arraybackslash}p{1cm}  
  >{\RaggedRight\arraybackslash}p{1.5cm}  
  >{\RaggedRight\arraybackslash}p{1.2cm}  
  >{\RaggedRight\arraybackslash}p{1.2cm}  
  >{\RaggedRight\arraybackslash}p{3cm}    
  >{\RaggedRight\arraybackslash}p{1.5cm}  
}
\toprule
\thead[l]{Dataset} &
\thead[l]{Viewing\\Distance (m)} &
\thead[l]{Viewing\\Target} &
\thead[l]{Setting} &
\thead[l]{Task type} &
\thead[l]{Distance\\Estimation} &
\thead[l]{Participants} &
\thead[l]{Collected Data Type} &
\thead[l]{Device} \\
\midrule

Funes Mora et al.~\cite{funes2014eyediap} &
R: up to 1.2 &
Monitor screen + 3D floating target &
Indoor &
Controlled &
X &
16 &
Eye-tracking, Viewing distance, Head pose, Face video &
RGB--D camera \\
\addlinespace[2pt]

Mansouryar et al.~\cite{mansouryar20163d} &
D: 1/1.25/1.5/1.75/2 &
Grid-based targets &
Indoor &
Controlled &
X &
14 &
Eye-tracking, Viewing distance, Eye video, Scene video &
PUPIL head-mounted eye tracker \\
\addlinespace[2pt]

Kothari et al.~\cite{kothari2020gaze, kothari2020processdata} &
R: No restrictions &
Various objects &
Indoor &
Real-world scenarios (e.g., ball catching) &
X &
19 &
Eye-tracking, Viewing distance, IMU, Scene video &
Pupil Labs ETG \\
\addlinespace[2pt]

Emery et al.~\cite{emery2021openneeds, emery2021openneeds_dataset} &
R: 0.3--5 &
Various objects &
VR & 
Real-world scenarios (e.g., drawing) &
X &
44 &
Eye-tracking, Viewing distance, IMU, Scene video, Hand pose (6DoF) &
Custom prototype VR-HMD \\
\addlinespace[2pt]

Koch et al.~\cite{koch2024deep, koch2024deepgaze_dataset} &
D: 0.5/1.5/3 &
Size-varied real and virtual targets &
AR & 
Controlled &
X &
4 &
Eye-tracking, Viewing distance, Scene video &
ARETT \\
\addlinespace[2pt]

Lee et al.~\cite{lee2017estimating} &
D: 1/2/3/4/5 &
Circle on a board &
Indoor &
Controlled &
O &
13 &
Eye-tracking, Viewing distance &
Pupil-labs eye tracker \\

\addlinespace[2pt]
C Elmadjian et al.~\cite{elmadjian20183d, elmadjian2018_3dgaze_dataset} &
D: 0.75/1.25/1.75\newline /2.25/2.75 &
Grid-based markers &
Indoor &
Controlled &
O &
11 &
Eye-tracking, Viewing distance &
Pupil Labs binocular head-mounted eye tracker \\
\addlinespace[2pt]

Liu et al.~\cite{liu20203d} &
D: 1/2/3/4 &
Patterned circular targets &
Indoor &
Controlled &
O &
10 &
Eye-tracking, Viewing distance, Depth map &
Intel Real Sense D435 RGB-D camera + two IR eye cameras \\
\addlinespace[2pt]

Walter~\cite{walter2022evaluation} &
D: 0.3/0.9/1.2/2 &
2D plane with the number &
XR & 
Controlled &
O &
10 &
Eye-tracking, Viewing distance, Head pose and rotation &
Varjo XR-3 \\
\addlinespace[2pt]

Stone et al.~\cite{stone2023gaze, stone2023gaze_dataset} &
R: 1.9--6.4 &
Circular targets throughout a room &
Indoor &
Controlled &
O &
10 &
Eye-tracking, Viewing distance &
Pupil Invisible \\
\addlinespace[2pt]

Cho et al.~\cite{cho2024hybrid} &
D: 0.5/1/1.3 &
Marker on a board &
Indoor &
Controlled &
O &
12 &
Eye-tracking, Viewing distance, Depth map &
Pupil-labs eye tracker \\
\addlinespace[2pt]

Hosp et al.~\cite{hosp2024simulation} &
D: 0.35/1/6 &
Virtual phone, Monitor, TV &
VR & 
Controlled &
O &
23 &
Eye-tracking, Viewing distance &
XTAL VR headset \\
\addlinespace[2pt]

von Behren et al.~\cite{von2025cnn, vonbehren2025gazedistance_dataset} &
R: 0.35--15 &
300 virtual targets &
VR & 
Controlled &
O &
41 &
Eye-tracking, Viewing distance, Depth map &
HTC Vive Pro Eye HMD \\
\addlinespace[2pt]

\midrule
GazeDepth (Ours) &
D: 0.33/0.5/3\newline
R: task-dependent ranges &
Various objects &
Indoor / Outdoor & 
Real-world scenarios &
O &
19 &
Eye-tracking, Viewing distance, IMU, Scene video &
Pupil Labs Neon eye tracker \\

\bottomrule
\end{tabularx}
\label{tab:distance_datasets}
\end{table}

%% file: Figure/fig_study_protocol.tex
\begin{figure}[t]
\centering
\includegraphics[width=0.75\textwidth]{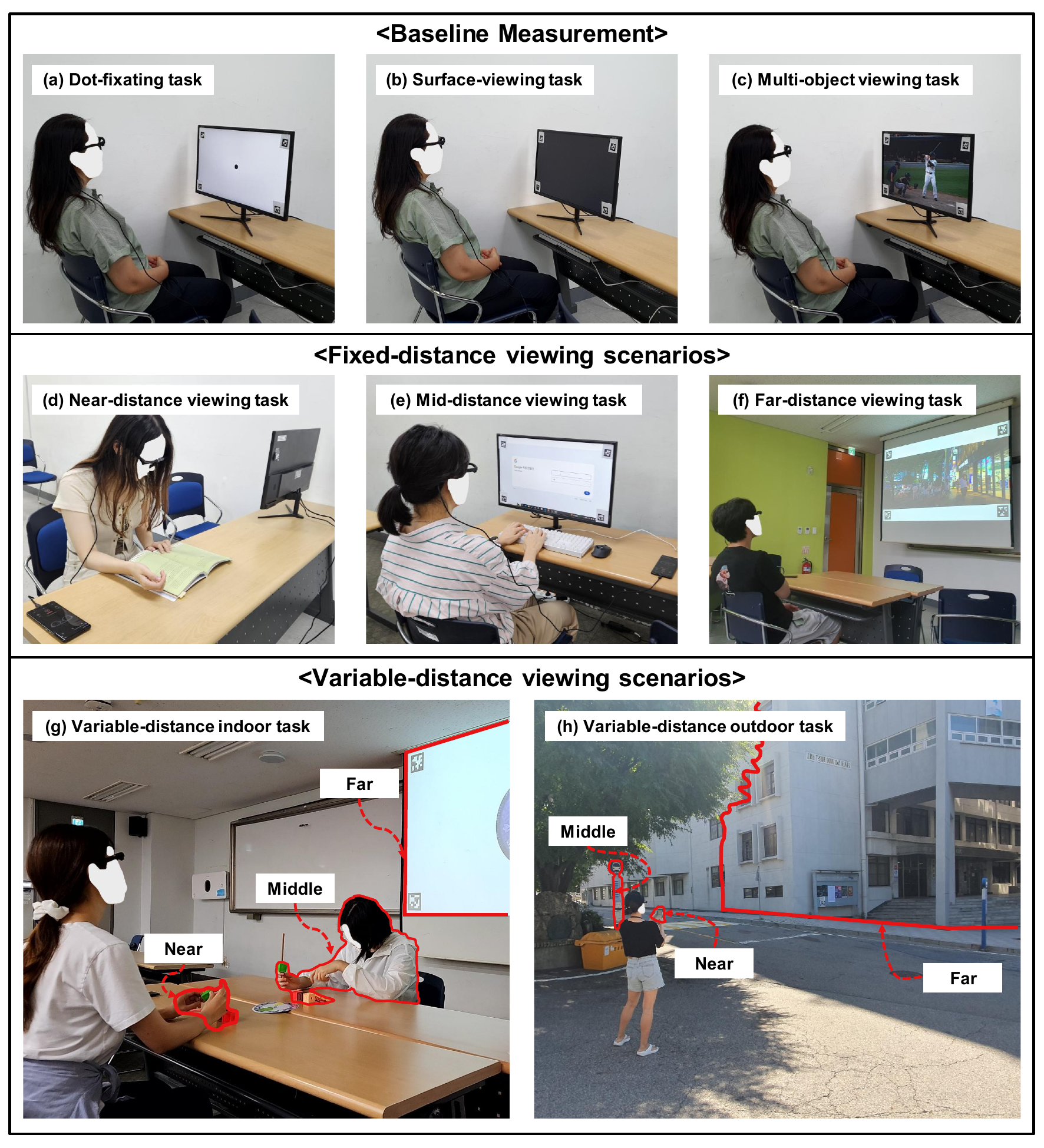}
\caption{Overview of the experimental protocol in the \dataset dataset. 
The study consists of three phases: (1) baseline measurement (dot-fixating, surface-viewing, and multi-object viewing), 
(2) fixed-distance viewing (near-, middle-, and far-distance viewing tasks), and 
(3) variable-distance viewing in indoor and outdoor environments.}
\label{fig:figure1}
\end{figure}

%% file: Table/tab_demography.tex
\begin{table}[t]
\centering
\caption{Participant demographics and ophthalmic characteristics (N = 19).}
\label{tab:participant_demographics}
\begin{tabular}{ll}
\toprule
\textbf{Category} & \textbf{Distribution} \\
\midrule
Gender & Male: 7, Female: 12 \\
Age (years) & Mean = 39.21, SD = 11.83 \\
Routine use of glasses/contact lenses & Yes: 7, No: 12 \\
History of ocular surgery & LASIK: 6, LASEK: 1, Lens implantation: 1 \\
Ocular disease status & Myopia only: 2, Hyperopia only: 1, \\
& Myopia + Astigmatism: 2, Presbyopia + Astigmatism: 1 \\
\bottomrule
\end{tabular}
\end{table}

%% file: Figure/fig_setup.tex
\begin{figure}[t]
\centering
\includegraphics[width=1\textwidth]{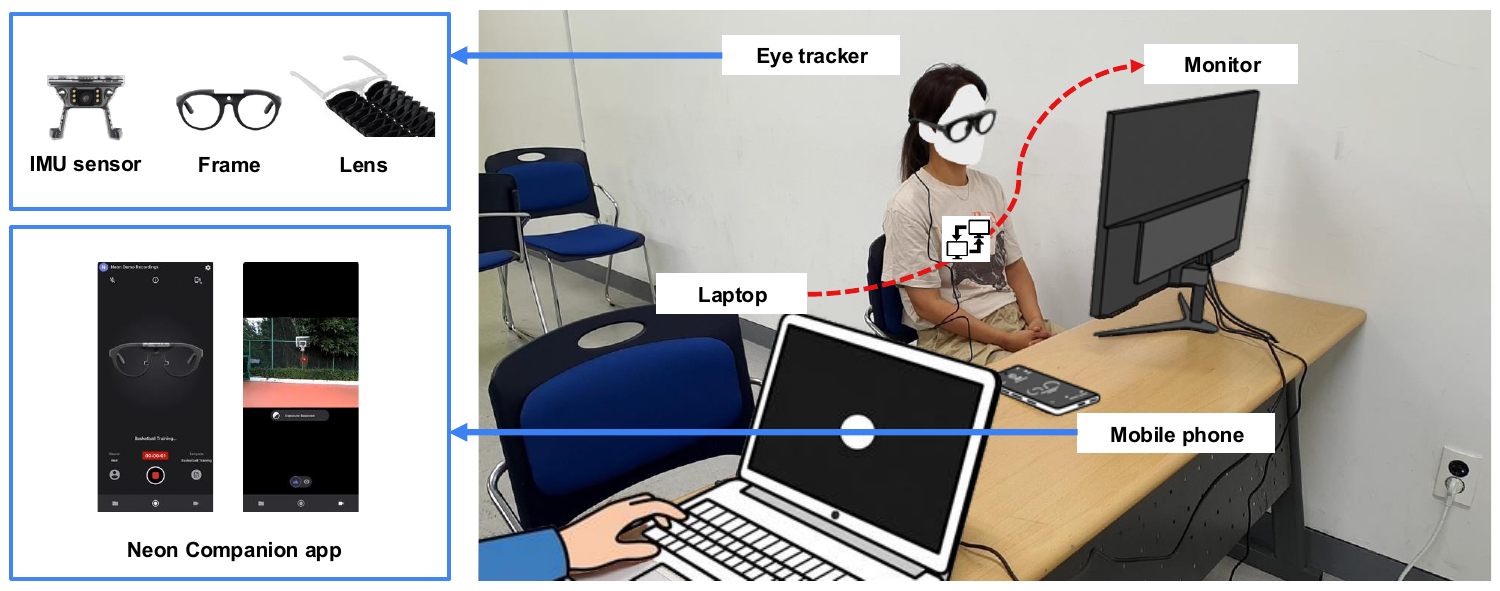}
\caption{Experimental setup used for data collection in the study. 
The figure illustrates the device configuration (e.g., eye-tracking), participant positioning, and the task operation workflow during recording sessions.}
\label{fig:setup}
\end{figure}

%% file: Figure/fig_auto_labeling_indoor_sample.tex
\begin{figure}[t]
\centering
\includegraphics[width=0.7\textwidth]{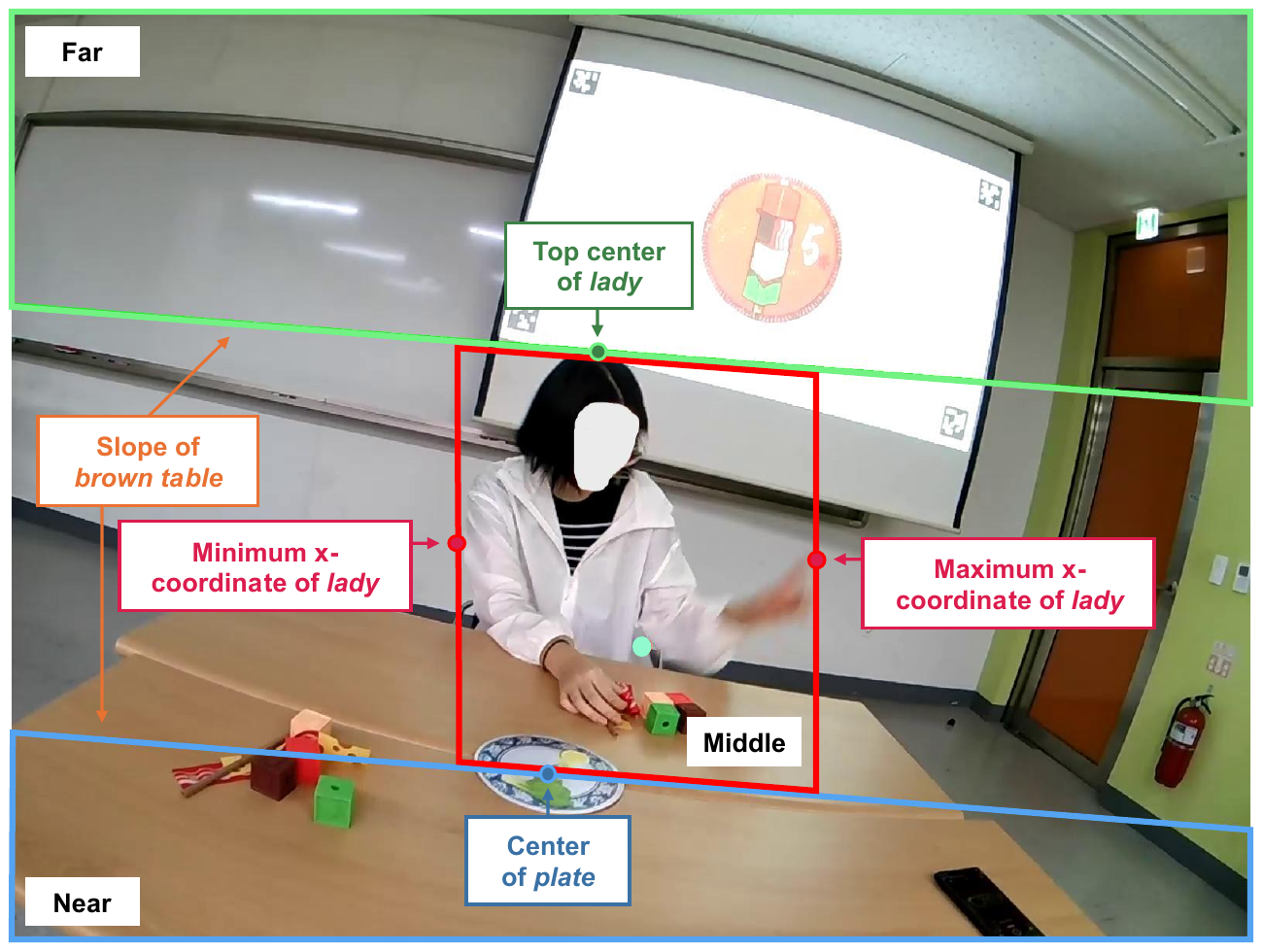}
\caption{Example of automatic annotation on an indoor sample frame. The image is partitioned into near (blue), middle (red), and far (green) regions. The gaze point is represented by a neon mint dot; since it is located within the middle region in this frame, the sample is assigned the middle label.}
\label{fig:auto_labeling_indoor_sample}
\end{figure}

%% file: Table/tab_auto_scenario.tex
\begin{table*}[t]
\centering
\caption{Cases for automatic labeling in the variable-distance indoor task were defined by the detection status of the reference objects used for zone definition (\textit{plate} and \textit{lady}) across all frames from all participants. The table also reports frame counts and proportions before and after \textit{plate}-position estimation.}
\label{tab:case_definition_before_after}
\small
\setlength{\tabcolsep}{5pt}
\begin{tabular}{c c c c c c c c c c}
\toprule
\multirow{2}{*}{\makecell[c]{\textbf{Case}\\\textbf{ID}}} &
\multicolumn{2}{c}{\makecell[c]{\textbf{Detection}\\\textbf{result}}} &
\multicolumn{3}{c}{\textbf{Labeling}} &
\multicolumn{2}{c}{\makecell[c]{\textbf{Before plate-position}\\\textbf{prediction}}} &
\multicolumn{2}{c}{\makecell[c]{\textbf{After plate-position}\\\textbf{prediction}}} \\
\cmidrule(lr){2-3}\cmidrule(lr){4-6}\cmidrule(lr){7-8}\cmidrule(lr){9-10}
& \textbf{Plate} & \textbf{Lady}
& \makecell[c]{\textbf{Near}\\\textbf{zone}}
& \makecell[c]{\textbf{Middle}\\\textbf{zone}}
& \makecell[c]{\textbf{Far}\\\textbf{zone}}
& \makecell[c]{\textbf{Frame}\\\textbf{count}}
& \makecell[c]{\textbf{Case}\\\textbf{ratio}}
& \makecell[c]{\textbf{Frame}\\\textbf{count}}
& \makecell[c]{\textbf{Case}\\\textbf{ratio}} \\
\midrule
1 & O & O & near & middle & far
  & 371,268 & 81.36\% & 390,181 & 85.51\% \\
2 & X & O & - & middle & far
  & 80,917 & 17.73\% & 62,004 & 13.59\% \\
3 & O & X & near & - & -
  & 4,023 & 0.88\% & 4,023 & 0.88\% \\
4 & X & X & - & - & -
  & 117 & 0.03\% & 117 & 0.03\% \\
\bottomrule
\end{tabular}

\vspace{0.4em}
\begin{minipage}{0.98\textwidth}
\footnotesize
\textbf{Case 1}: Scene divided into near, middle, and far zones.
\textbf{Case 2}: Scene divided into middle and far zones.
\textbf{Case 3}: The scene includes the near zone and an undefined upper region, or the \textit{lady} is within view but not detected by the model.
\textbf{Case 4}: The participant either tilted their head markedly upward, or eye-tracker tracking was disturbed.
\end{minipage}
\end{table*}

%% file: Figure/fig_annotator_tool.tex





\begin{figure}[t]
\centering
\includegraphics[width=1\textwidth]{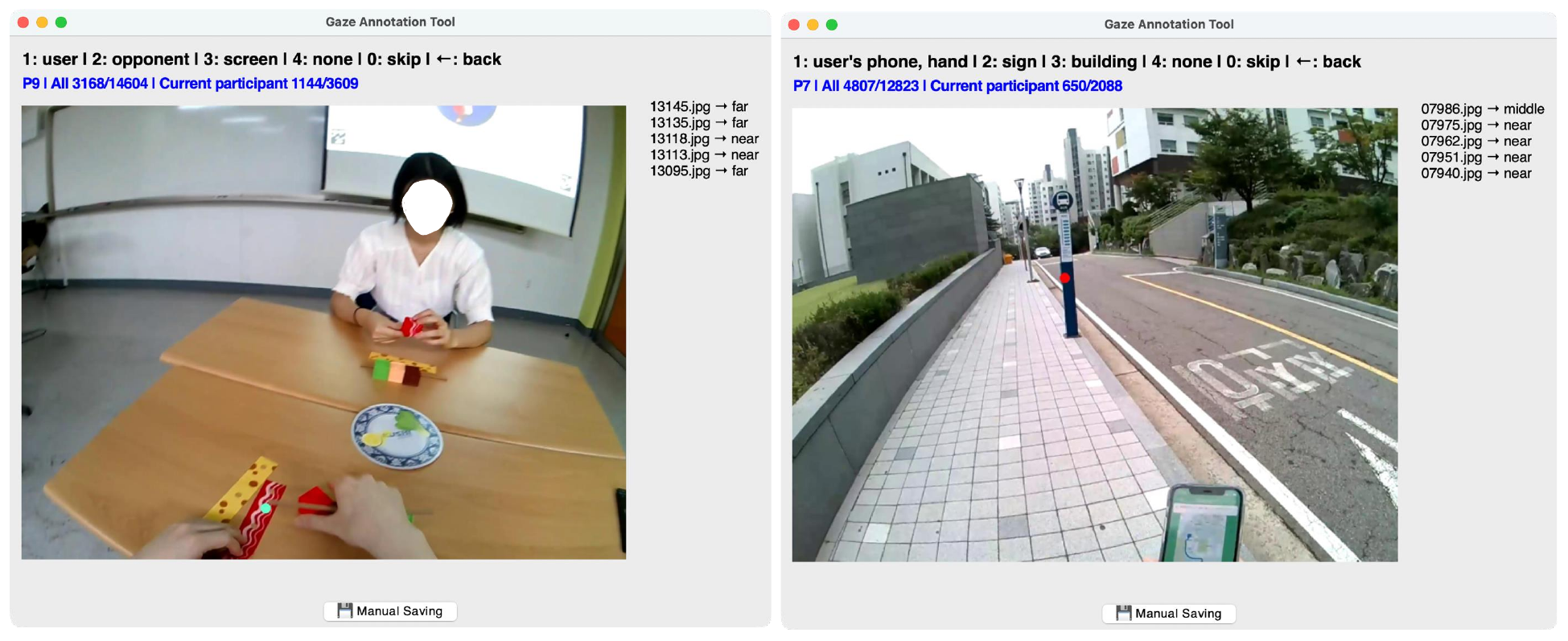}
\caption{Annotation tool interfaces used for gaze distance manual labeling in variable-distance indoor and outdoor tasks. 
The tool displays each frame with its gaze point and supports keyboard-based class assignment (1:near /2:middle /3:far /4:none /0:skip).}
\label{fig:annotator_tool_indoor_outdoor}
\end{figure}

%% file: Figure/fig_boxplot.tex
\begin{figure}[t]
  \centering
  \includegraphics[width=\linewidth]{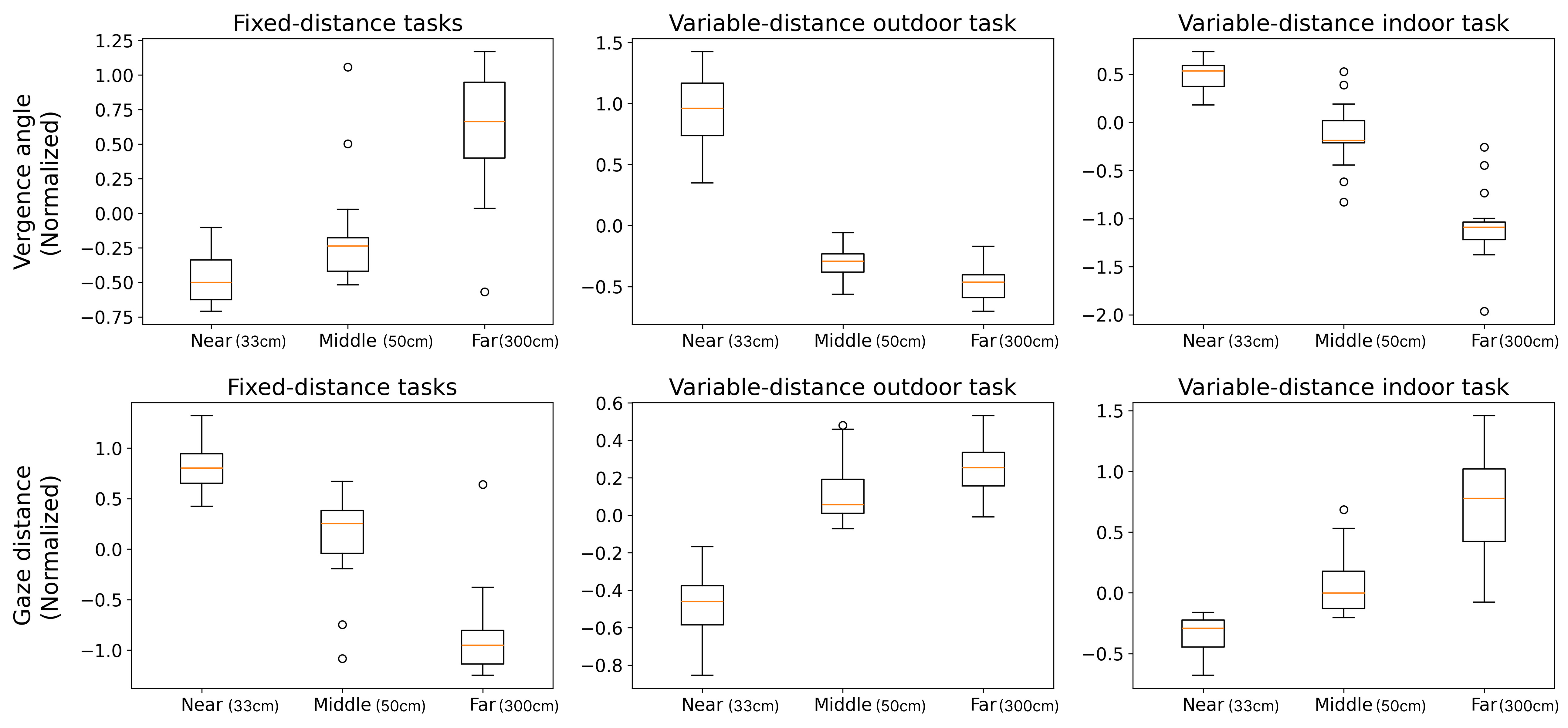}
  \caption{Distribution of vergence angle (top row) and vergence-based gaze distance (bottom row) across viewing-distance conditions (near/middle/far) in the fixed-distance task and the variable-distance indoor/outdoor tasks. Values are participant-normalized (z-scored).}
  \label{fig:distance_validation_boxplots}
\end{figure}

%% file: Table/tab_modelResult.tex
\begin{table*}[t]
\centering
\caption{Classification performance for Near/Middle/Far distance conditions across three tasks (fixed-distance viewing, variable-distance indoor, and variable-distance outdoor). Performance was evaluated using accuracy and F1-score across repeated runs under three window-size settings. Values are reported as Mean (SD).}
\label{tab:model_performance_all}
\begin{tabular}{llcccccc}
\toprule
\multirow{2}{*}{$W$} & \multirow{2}{*}{Model}
& \multicolumn{2}{c}{Fixed-distance viewing}
& \multicolumn{2}{c}{Variable-distance indoor}
& \multicolumn{2}{c}{Variable-distance outdoor} \\
\cmidrule(lr){3-4} \cmidrule(lr){5-6} \cmidrule(lr){7-8}
&& Acc & F1 & Acc & F1 & Acc & F1 \\
\midrule

\multirow{6}{*}{$0.5s$} 
& MLP &  \underline{0.904 (0.06)} & \underline{0.905 (0.06)} & 0.796 (0.08) & 0.741 (0.06) & 0.723 (0.05) & 0.656 (0.04) \\
& Decision Tree &  0.820 (0.09) & 0.820 (0.09) & 0.746 (0.05) & 0.681 (0.04) & 0.661 (0.03) & 0.615 (0.02) \\
& Random Forest &  0.879 (0.10) & 0.879 (0.10) & \textbf{0.827 (0.05)} & \textbf{0.742 (0.05)} & \textbf{0.763 (0.04)} & \textbf{0.662 (0.04)} \\
& kNN  &  0.850 (0.08) & 0.850 (0.08) & 0.784 (0.05) & 0.689 (0.04) & 0.719 (0.04) & 0.641 (0.03) \\
& LDA  &  \textbf{0.919 (0.04)} & \textbf{0.920 (0.04)} & 0.819 (0.05) & \underline{0.736 (0.05)} & 0.752 (0.04) & \underline{0.656 (0.04)} \\
& XGBoost  &  0.886 (0.10) & 0.886 (0.10) & \underline{0.822 (0.06)} & 0.724 (0.06) & \underline{0.760 (0.04)} & 0.649 (0.04) \\
\cmidrule(lr){2-8}
& Majority Voting  & 0.325 (0.01) & 0.163 (0.00) & 0.562 (0.09) & 0.238 (0.02) & 0.420 (0.05) & 0.196 (0.01) \\

\midrule

\multirow{6}{*}{$1s$} 
& MLP           & \underline{0.914 (0.06)} & \underline{0.915 (0.06)} & 0.786 (0.07) & \textbf{0.724 (0.06)} & 0.702 (0.04) &  0.642 (0.04) \\
& Decision Tree & 0.828 (0.10) & 0.827 (0.10) & 0.730 (0.05) & 0.660 (0.05) & 0.649 (0.04) & 0.601 (0.03) \\
& Random Forest & 0.894 (0.10) & 0.894 (0.10) & \textbf{0.814 (0.05)} & \underline{0.712 (0.05)} & \textbf{0.756 (0.04)} & \textbf{0.653 (0.04)} \\
& kNN           & 0.846 (0.08)  & 0.846 (0.08)  & 0.756 (0.05) & 0.639 (0.04) & 0.700 (0.04) & 0.618 (0.03) \\
& LDA           & \textbf{0.931 (0.04)}  & \textbf{0.932 (0.04)}  & 0.804 (0.05) & \underline{0.712 (0.04)} & 0.740 (0.04) & \underline{0.644 (0.04)} \\
& XGBoost       & 0.899 (0.10)  & 0.899 (0.10)  & \underline{0.811 (0.05)} & 0.703 (0.06) & \underline{0.752 (0.04)} & \underline{0.644 (0.04)} \\
\cmidrule(lr){2-8}
& Majority Voting  &  0.325 (0.01) & 0.163 (0.00) & 0.562 (0.09) & 0.238 (0.02) & 0.426 (0.05) & 0.198 (0.01) \\

\midrule

\multirow{6}{*}{$10s$} 
& MLP           &  \underline{0.947 (0.05)} & \underline{0.945 (0.05)} & 0.764 (0.08) & \textbf{0.601 (0.07)} & 0.688 (0.07) & 0.620 (0.06) \\
& Decision Tree &  0.886 (0.08) & 0.884 (0.08) & 0.698 (0.07) & 0.532 (0.07) & 0.626 (0.07) & 0.551 (0.06) \\
& Random Forest &  0.942 (0.07) & 0.940 (0.08) & \textbf{0.799 (0.07)} & 0.556 (0.05) & \textbf{0.744 (0.05)} & 0.620 (0.07) \\
& kNN           &  0.777 (0.08) & 0.769 (0.09) & 0.678 (0.09) & 0.381 (0.05) & 0.571 (0.05) & 0.448 (0.05) \\
& LDA           &  \textbf{0.966 (0.04)} & \textbf{0.966 (0.04)} & 0.772 (0.09) & \underline{0.587 (0.06)} & 0.736 (0.06) & \textbf{0.644 (0.05)} \\
& XGBoost       &  0.949 (0.05) & 0.948 (0.05) & \underline{0.795 (0.07)} & \underline{0.587 (0.08)} & \underline{0.741 (0.06)} & \underline{0.631 (0.08)} \\
\cmidrule(lr){2-8}
& Majority Voting  &  0.321 (0.02) & 0.161 (0.01) & 0.670 (0.12) & 0.265 (0.02) & 0.430 (0.10) & 0.197 (0.03) \\

\bottomrule
\end{tabular}
\end{table*}

%% file: Figure/SHAP_all.tex
\begin{figure*}[t]
    \centering
    \includegraphics[width=\textwidth]{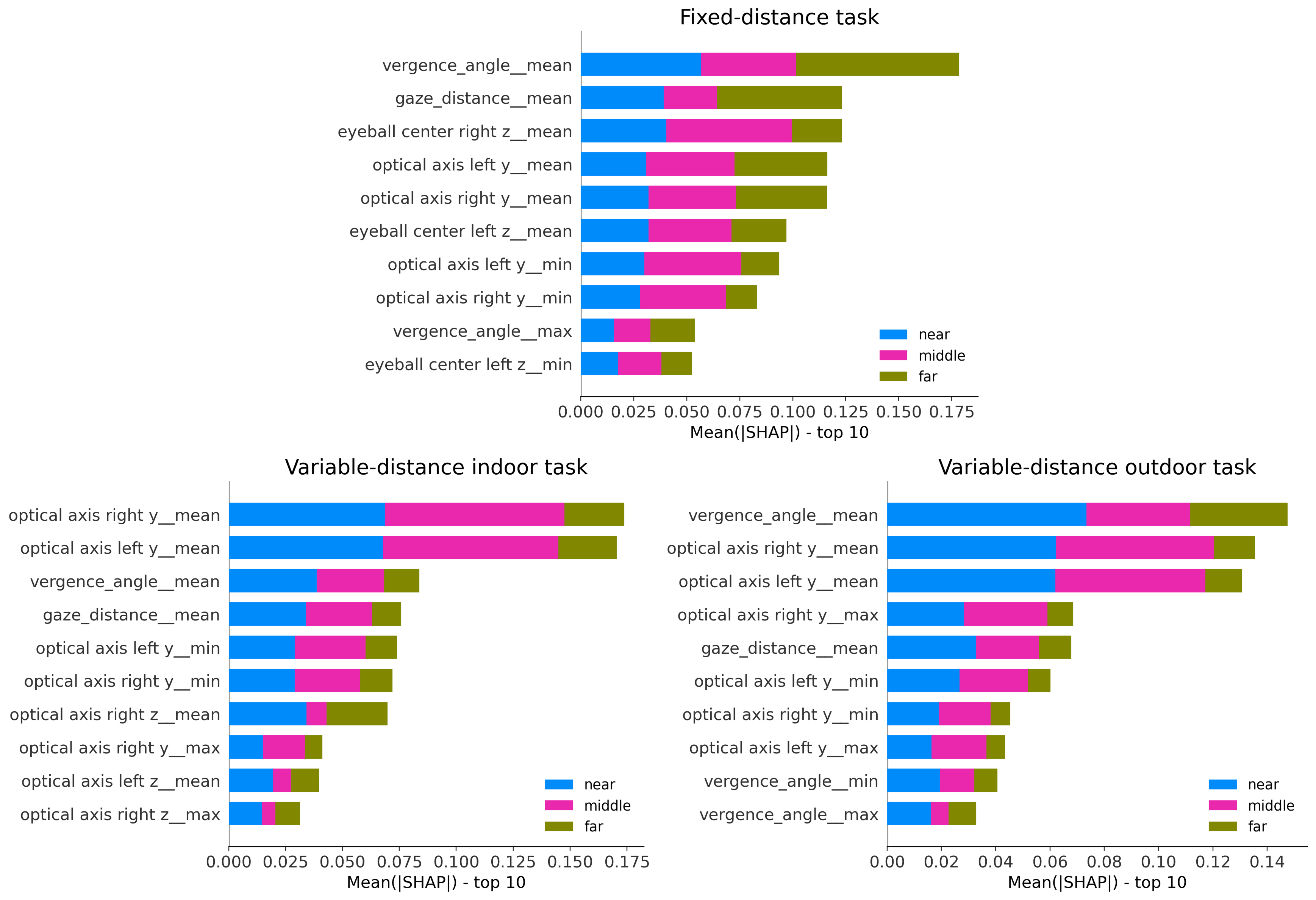}
    \caption{The Top 10 Features of SHapley Additive ex-Planations (SHAP) for the Random Forest model across three tasks: fixed-distance viewing (top), variable-distance indoor viewing (bottom left), and variable-distance outdoor viewing (bottom right). Bars indicate mean absolute SHAP values, decomposed by class (near/middle/far).}
    \label{fig:shap_all}
\end{figure*}